\documentstyle[graphicx,aps]{revtex}
\begin{document}
\twocolumn
\draft
{\bf Reply on T\~oke's and Schr\"oder's comment on our paper ``Statistical
nature of multifragmentation''}
\cite{gross165}. T\~oke and Schr\"oder comment \cite{toke98} on the following
points: 
\begin{enumerate}
\item the wrong citation in \cite{toke97} of {\em MMMC} as the Berlin SMM
model: ``is obviously irrelevant''.
\item the use of the binomial fragment distribution in \cite{toke97} at
small excitation energies : ``This is irrelevant as singularities in the
theoretical Pseudo Arrhenius plot occur at rather high excitation
energies. It is more than debatable whether the approximately $400$MeV
excitation energy, where the calculations of \cite{gross165} begin,
qualify as very small excitation when the fragment production is
forbidden by energy conservation.''
\item Disregard the effect of the imperfections of the experimental
setup (experimental filter) : ``..irrelevant..as the used filter in 
\cite{gross165} is too strong...the excess $E_t$  in fig.1 of 
\cite{gross165} due to an improper inclusion of fission fragments in
the definition of $E_t$, contrary to the experimental definition.''
\end{enumerate}
1) We agree with statement $1$, but we think a more carefull reading
and quotation of the literature is certainly usefull as it
demonstrates how far the authors understand (or not) the theoretical
models and their differences, see below.\\ 2) At excitation energy of
$2-3$MeV/nucleon the multifragmentation channels opens, consequently
the binomial distribution which ignores this threshold and evidently
violates energy conservation is not adequate.\\ 3) Without a
discussion of the experimental filter in this context makes an
experimental paper like ref.\cite{toke97} useless for a theoretical
interpretation. Of course fission events are there and are suppressed
not by the fragmentation mechanism itself but by the detection method
(filter). Therefore, our procedure to include it and filter the
fission events out at the end seems to us to be the correct
procedure.\\ The discussion of ``retrieval of information lost'' is
vague and makes little sense. We've shown how the filter influences
the observed mean values and variances. Of course can the pole in the
primary $E_t$ distribution be filtered away.

The following criticism by T\~oke and Schr\"oder on {\em MMMC}
not to be microcanonical must be due to a misunderstanding:
As explained in any detail in both reviews
\cite{gross95,gross153} {\em MMMC} observes strictly overall energy 
conservation and treats every exit quantum state, characterized by a
{\em complete} set of quantum numbers with the same weight. The
momenta of the neutrons are integrated over under the constrain of
fixed total momentum (cm-momentum) and fixed total kinetic energy
within their common cm-system. Therefore, the weight of all neutrons
{\em together} is the volume of all neutrons phase space. This is done
separately for the prompt as well for the evaporated neutrons, which are
treated as unbound resonances in the s.p.potential of the fragments. In that
respect and some others {\em MMMC} is different from the Copenhagen SMM model
which is semi-microcanonically switches for each selected partition
into the canonical ensemble considering neutron evaporation mainly as
a slow secondary process (this is discussed in detail in
\cite{gross125} and also by the Copenhagen group in
\cite{bondorf98,bondorf95}).\\ {\em A.Botvina and D.H.E.Gross,\\
Hahn-Meitner Institut,Glienickerstr.100,\\D14109 Berlin}


\end{document}